\documentclass[fleqn,usenatbib]{mnras}
\usepackage[utf8]{inputenc}
\usepackage{newtxtext,newtxmath}
\usepackage[T1]{fontenc}
\usepackage{graphicx}
\usepackage{amsmath}

\title[Characterisation of TIC 60040774]{Characterisation of the eclipsing post-common-envelope binary TIC 60040774}
\author[R. Priyatikanto et al.]{
R. Priyatikanto,$^{1}$\thanks{E-mail: rhorom.priyatikanto@brin.go.id (RP)}
C. Knigge,$^{2}$
S. Scaringi,$^{3}$
J. Brink$^{4,5}$
and D.A.H. Buckley$^{4,5,6}$
\\
$^{1}$Research Center for Space, National Research and Innovation Agency, Bandung 40132, Indonesia\\
$^{2}$School of Physics and Astronomy, University of Southampton, Highfield, Southampton SO17 1BJ, United Kingdom\\
$^{3}$Centre for Extragalactic Astronomy, Department of Physics, Durham University, DH1 3LE, United Kingdom\\
$^{4}$South African Astronomical Observatory, PO Box 9, Observatory, 7935, Cape Town, South Africa\\
$^{5}$Department of Astronomy, University of Cape Town, Private Bag X3, Rondebosch 7701, South Africa\\
$^{6}$Department of Physics, University of the Free State, PO Box 339, Bloemfontein 9300, South Africa\\
}

\date{Accepted XXX. Received YYY; in original form ZZZ}

\pubyear{2022}

\newcommand{\tmag}{T_{\text{mag}}}
\newcommand{\teff}[1]{T_{\text{eff{#1}}}}
\newcommand{\logg}[1]{\log(g_{#1})}

\begin{document}
\label{firstpage}
\pagerange{\pageref{firstpage}--\pageref{lastpage}}
\maketitle

\begin{abstract}
Binaries with a white dwarf primary and a main sequence secondary can be used to test our understanding of both single and binary star evolution. A small fraction of such systems experienced a common-envelope phase from which they emerged with a relatively short orbital period. Here, we present the characterisation of an eclipsing post-common-envelope binary of this kind, TIC 60040774, based on the light curve provided by the Transiting Exoplanet Survey Satellite (TESS), multi-band photometry collated from the virtual observatory, and spectroscopic data obtained the Southern African Large Telescope (SALT). With an orbital period of $0.404807\pm0.000149$ days, this system consists of a young white dwarf paired with an M6.5 dwarf companion. We estimate the masses of the primary and secondary to be $0.598\pm0.029$ M$_{\odot}$ and $0.107\pm0.020$ M$_{\odot}$, while the effective temperatures are $14050\pm360$ K and $2759\pm50$ K, respectively. The eclipse ingress and egress profile is shallower than expected from a simple geometric model such that more precise high-cadence photometry is required to understand the nature of this system. Given the similarity of TIC 60040774 to systems like GK Vir and NN Ser, it will be worth tracking its eclipse times to check for the presence of one or more circumbinary planets. 
\end{abstract}


\section{Introduction}
Binary systems can provide important insights into stellar structure and evolution. Binaries, and especially eclipsing binaries, allow us to determine fundamental stellar parameters (such as masses and radii) and to study their evolution as a function of stellar age\citep{andersen1991, torres2010}.  Binary stars are therefore central to our understanding of stellar evolution \citep{stancliffe2015, claret2018}. Eclipsing binaries also provide opportunities to determine distances on extra-galactic scales with great precision \citep{bonanos2006, vilardell2010, pietrzynski2013}, helping to anchor the cosmological distance ladder. 

Binaries are not rare amongst stars in the Milky Way. \cite{moe2017} estimate that almost all massive stars tend to be in binary or multiple systems, with a multiplicity fraction of 90\%. This fraction decreases roughly linearly with the logarithm of the primary mass, so, for example, still ${\gtrsim}40\%$ of solar-like stars belong to multiple systems. {\em Close} binaries are also quite common. For example, ${\sim}15\%$ of solar-like stars become the primary of close binary systems with separation of less than ${\sim}13$ AU \cite{moe2017}. These systems will become interacting binaries where the evolution of the components are affected by mass transfer either through Roche-lobe overflow or a common-envelope phase.  These binary evolution channels give many exotic objects, such as blue stragglers \citep{perets2009}, cataclysmic variables \citep{knigge2011} and AM CVn stars \citep[e.g.,][]{nelemans2004}. Common-envelope evolution, in particular, also produces a significant population of close, but detached  white-dwarf/main-sequence (WDMS) binaries.

\citet{rebassa2021} recently estimated a WDMS space density of $3.7\times10^{-4}$ pc$^{-3}$, implying the presence of $\simeq 1500$ such systems within a distance of $100$~pc. Approximately a quarter of WDMS are close enough that they must be post-common-envelope binaries (PCEBs), and $\simeq 10\%$ of these PCEBs exhibit eclipses \citep{rebassa2016, parsons2013, parsons2015}. 

Eclipsing WDMS are essential objects for constraining the mass-radius relationships for white dwarfs \citep[WDs][]{parsons2017} and low-mass main-sequence (MS) stars \citep{parsons2018}. Moreover, multi-epoch observations of some eclipsing PCEB systems reveal cyclic variations in the eclipse timings that cannot be ascribed to magnetic activity via the  Applegate mechanism \citep{applegate1992}. Examples here include NN Ser \citep{beuerman2010, marsh2014}, RR Cae \citep{qian2012}, DE CVn \citep{han2018}, and GK Vir \cite{almeida2020}. In these systems, the eclipse time variations are therefore thought to be due to the presence of distant sub-stellar or planetary companions. The existence of such companions to PCEBs raises obvious questions for the theory of planetary formation \citep{bear2014, volschow2014}. P-type or circumbinary planets that were formed from the same cloud as the binary itself are expected to be ejected during the common-envelope phase, due to the abrupt loss of gravitational binding energy. On the other hand, S-type planets initially bound to one of the components component are expected to be accreted by the primary, producing metal pollution signatures in the white dwarf's atmosphere \citep{koester2014}. Consequently, post-common-envelope planetary formation has been proposed to explain the existence of  planetary bodies orbiting systems like NN Ser \citep{schleicher2014, volschow2014}. In order to test such scenarios, it is vital to determine the frequency and statistical properties of sub-stellar and planetary companions among a larger sample of eclipsing PCEBs. 

In this paper, we present the discovery and characterisation of a bright new eclipsing PCEB. The system, TIC~60040774, is a high-proper-motion object that lies close to the WD sequence in the colour-magnitude diagram and is located fairly nearby, at a distance of $134$ pc. We show that TIC~60040774 is close binary in which an M6.5V secondary eclipses a WD primary every $\simeq 0.4$~days. We also determine the binary parameters of the system and discuss its properties in the context of other PCEBs.

\section{Data}
TIC 60040774 ($G=16.98$) is located ${\sim}15^{\circ}$ south of the Galactic plane. Based on its {\em Gaia} EDR3 astrometry, it exhibits a relatively higher proper motion of $\simeq 68$~mas~yr$^{-1}$ and is located at a distance of $134.465$ parsecs \citep{edr3}. Some basic parameters of TIC 60040774 are summarised in Table \ref{tab:basic}. According to its location in the Hertzprung-Russell diagram (see Figure 1), \citet{gentile2021} identified this object as a highly probable white dwarf with an estimated mass of ${\sim}0.3$ M$_{\odot}$ (assuming it to be a single WD).  TIC~60040774 was observed by the Transiting Exoplanet Survey Satellite \citep[TESS,][]{ricker2015} with two-minute cadence and identified as a target with significant transit by the Science Processing Operations Center (SPOC) Pipeline \citep{caldwell2020}. However, the Data Validation pipeline, which fits limb-darkened transiting planet models to the observed light curve \citep{twicken2018}, excluded TIC 60040774 as a target of interest, since it assumed a solar-like primary and predicted the secondary to be a non-planetary body with an estimated radius of ${\sim}0.8$ R$_{\odot}$. \cite{priyatikanto2022} independently (re-)discovered TIC 60040774 as an eclipsing binary during a search for highly variable WDs in the \textit{TESS} data set. We then collected additional multi-wavelength photometric measurements from the literature and obtained new spectroscopic observation with the Southern African Large Telescope (SALT).

\begin{figure*}
    \centering
    \includegraphics[scale=0.8]{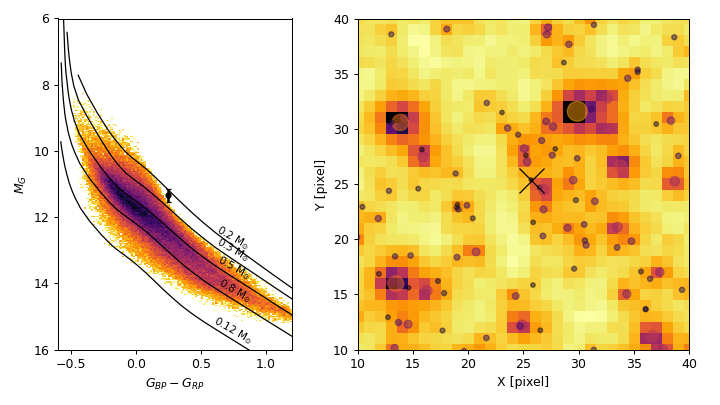}
    \caption{Left: the position of TIC 60040774 in the HR diagram plotted over the probable white dwarfs from \citet{gentile2021}. Black lines are the cooling tracks of DA white dwarfs with masses from $0.2$ to $1.5$ solar masses. Right: \textit{TESS} field-of-view around TIC 60040774 (cross) complemented with stars with $\tmag<16.5$.}
    \label{fig:fov}
\end{figure*}

\begin{table}
    \caption{Basic parameters of TIC 60040774}
    \label{tab:basic}
    \centering
    \begin{tabular}{l|l|l}
    \hline
    Other designations & \multicolumn{2}{p{0.5\columnwidth}}{Gaia DR2 2943496741862496256 \newline
    2MASS J06195643-1758186
    } \\
    \hline
    \\
    \textbf{Astrometry}\\
    \hline
    Right ascension & $\alpha_{2000}$ & $06^{\text{h}}\ 19^{\text{m}}\ 56.434^{\text{s}}$ \\ 
    Declination & $\delta_{2000}$ & $-17^{\circ}\ 58'\ 18.621''$ \\
    Galactic longitude & $l$ & $225.796191^{\circ}$ \\
    Galactic latitude & $b$ & $-14.872777^{\circ}$ \\
    Proper motion & $\mu_{\alpha}\cos\delta$ & $-23.306\pm0.052$ mas/yr \\
    & $\mu_{\delta}$ & $-63.937\pm0.057$ mas/yr \\
    Parallax & $\pi$ & $7.4369\pm0.0611$ mas \\
    Distance & $d$ & $134.465\pm1.105$ pc \\
    \hline
    \\
    \textbf{Photometry}\\
    \hline
    Gaia EDR3 & $G$ & $16.97695\pm0.00296$ \\
    & $B_P$ & $17.04543\pm0.00596$ \\
    & $R_P$ & $16.79837\pm0.00722$ \\
    TESS & $\tmag$ & $16.880\pm0.009$ \\
    Pan-STARRS1 & $g$ & $16.949\pm 0.004$\\
        & $r$ & $17.168\pm 0.003$\\
        & $i$ & $17.208\pm 0.003$\\
        & $z$ & $17.034\pm 0.006$\\
        & $y$ & $16.797\pm 0.009$\\
    2MASS & $J$ & $15.675\pm 0.054$\\
        & $H$ & $15.220\pm 0.074$\\
        & $Ks$ & $14.801\pm 0.099$\\
    AllWISE & $W1$ & $14.737\pm 0.035$\\
        & $W2$ & $14.611\pm 0.064$\\
    \hline
    \end{tabular}
\end{table}

\subsection{SALT spectra}
The Robert Stobie Spectrograph \citep[RSS,][]{burgh2003}, operated at the prime focus of SALT, was used to acquire optical spectra of TIC 60040774 on 2021-11-14 22:48 UT and 2021-11-16 22:19 UT. The observations were taken in long-slit spectroscopy mode and used the PG0900 grating, yielding a resolution of $R = \lambda/\Delta\lambda\approx 1065$ at the central wavelength of $6050$ \AA. The blue and red segments of the optical spectrum were acquired separately. For the blue segment ($3900$ to $7000$ {\AA}), a UV blocking filter was used, while the red segment ($6000$ to $9000$ \AA) was obtained with a blue blocking filter. Exposure times of $1200$ s yielded sufficient signal-to-noise for our purposes. PySALT \citep{crawford2010} was used to reduce and calibrate the spectra.

\subsection{Multiband photometry}
The characterisation of the faint, red secondary component requires access to wavelengths beyond the optical range. Fortunately, TIC 60040774 was observed by several survey missions, allowing the construction of a broad-band Spectral Energy Distribution (SED) that extends well into the infrared region. We used the Virtual Observatory SED Analyzer \citep[VOSA\footnote{\url{http://svo2.cab.inta-csic.es/theory/vosa/}},][]{bayo2008} to access and analyse these photometric data sets, specifically the {\em Gaia} Early Data Release 3 \citep[Gaia EDR3,][]{edr3}, Panoramic Survey Telescope and Rapid Response System Data Release 1 \citep[Pan-STARRS1,][]{chambers2016}, Two Micron All Sky Survey catalogue of point sources \citep[2MASS,][]{skrutskie2006}, and the AllWISE source catalogue \citep{cutri2014} from the Wide-field Infrared Survey Explorer. The observed magnitudes of TIC 60040774 from these catalogues are summarised in Table \ref{tab:basic}. Using specific zero points for selected photometric filters, those magnitudes were transformed into flux densities and then compared to the synthetic SEDs generated from selected models. The synthetic SEDs were normalised by adopting the distance of $d=134.465\pm1.105$ derived from {\em Gaia} EDR3 parallax \citep{edr3}. At this distance, the Galactic extinction in the direction of TIC 60040774 is negligible \citep{green2019}\footnote{\url{http://argonaut.skymaps.info/}}.

\subsection{TESS light curve}
\begin{figure}
    \centering
    \includegraphics[width=\columnwidth]{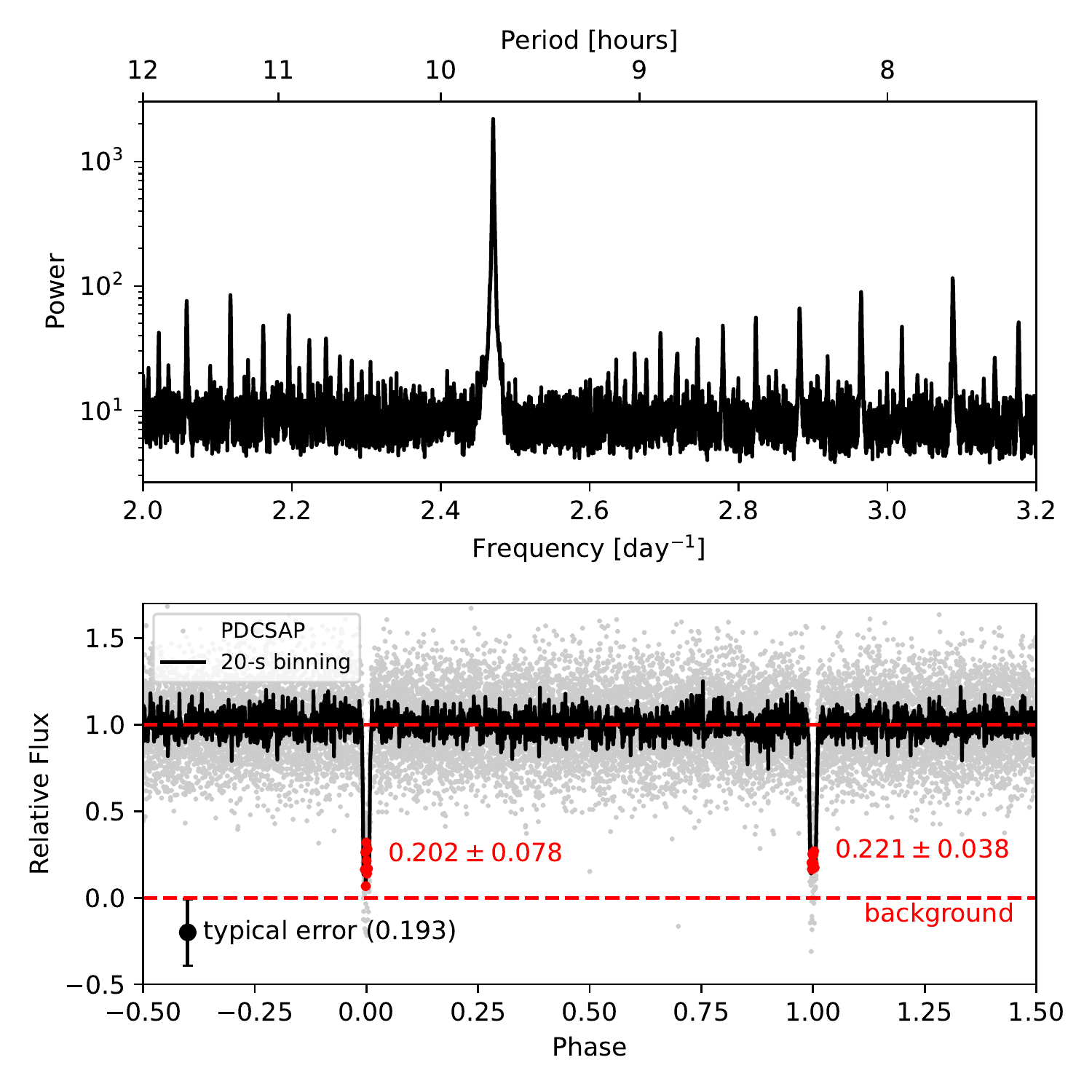}
    \caption{Top: BLS periodogram of TIC 60040774 shows prominent periodicity at $P=0.404807$ days. Bottom: folded light curve shows primary eclipse with very short duration. Normalised fluxes from PDCSAP are represented by grey dots with typical error of $0.193$. To suppress the scatter, the light curve is binned (shown as black line). There is no significant difference between even and odd eclipses as indicated by the mean and standard deviation of fluxes during the eclipses.}
    \label{fig:periodogram}
\end{figure}

TIC 60040774 was observed by \textit{TESS} in Sector 33 from 17 December 2020 to 13 January 2021. \textit{TESS} uses four cameras comprising 16 CCDs to monitor the variability of the sky in sectors that change every 27.4 days \citep{ricker2015}. Each sector covers an area of $24^{\circ}\times90^{\circ}$ with $21"$ pixel resolution and 2 s exposure time. Full-frame images are not stored at that cadence, however. Instead, different cadences are adopted, depending on the target priorities. There are 20 s cadence data products for approximately 1000 target stars for astroseismology, 120 s cadence data products for planet search priority targets, and 30 min cadence full-frame images for general-purpose photometry. Within this hierarchy, TIC~60040774 was observed with 120 s cadence. The data was processed using the \textit{TESS} Science Processing Operations Center pipeline to produce calibrated target pixels, simple aperture photometry flux time series (SAP), and pre-search data conditioning corrected flux time series (PDCSAP) \citep{caldwell2020}. The light curve for TIC 60040774 is available at the Mikulski Archive for Space Telescopes (MAST) and easily accessed and processed using \textsc{Lightkurve} Python package \citep{lightkurve2018}. Some analyses were also performed using \texttt{lightkurve} package, including the recovery of the eclipse period through the construction a box least square (BLS) periodogram \citep{kovacs2002}, which is suitable for analysing periodic transit-like events. Simple light curve models with periods ranging from $0.3$ to $0.5$ days and transit/eclipse durations from $1.44$ to $14.4$ minutes were evaluated to construct the periodogram shown in Figure \ref{fig:periodogram}). This yields an optimal period estimate of $P = 0.404807$~days.

As depicted on the right panel of Figure \ref{fig:fov}, the object is surrounded by several brighter stars that might contaminate the target flux within the specified aperture. There are four stars with $\tmag<16.5$ within $1^{\circ}$ radius from TIC 60040774, including TIC 60040779 and TIC 60040763, with magnitudes of $12.7$ and $13.9$, respectively. These are located less than 2 pixels from the target star. In a dense region, point spread function (PSF) modelling might be useful to remove the contamination from the neighboring stars \citep[e.g.][]{nardiello2019}. However, this is not required in our case, since the neighbouring stars show no variability. The flux time series data produced by the TESS-SPOC behaves very well, as indicated by the linearity of TIC 60040774 and some neighboring stars in flux-magnitude space. We confirmed that the PDCSAP flux is background-corrected and free from contamination.

\section{Characteristics of TIC 60040774}
\subsection{White dwarf parameters}

\begin{figure}
    \centering
    \includegraphics[scale=0.7]{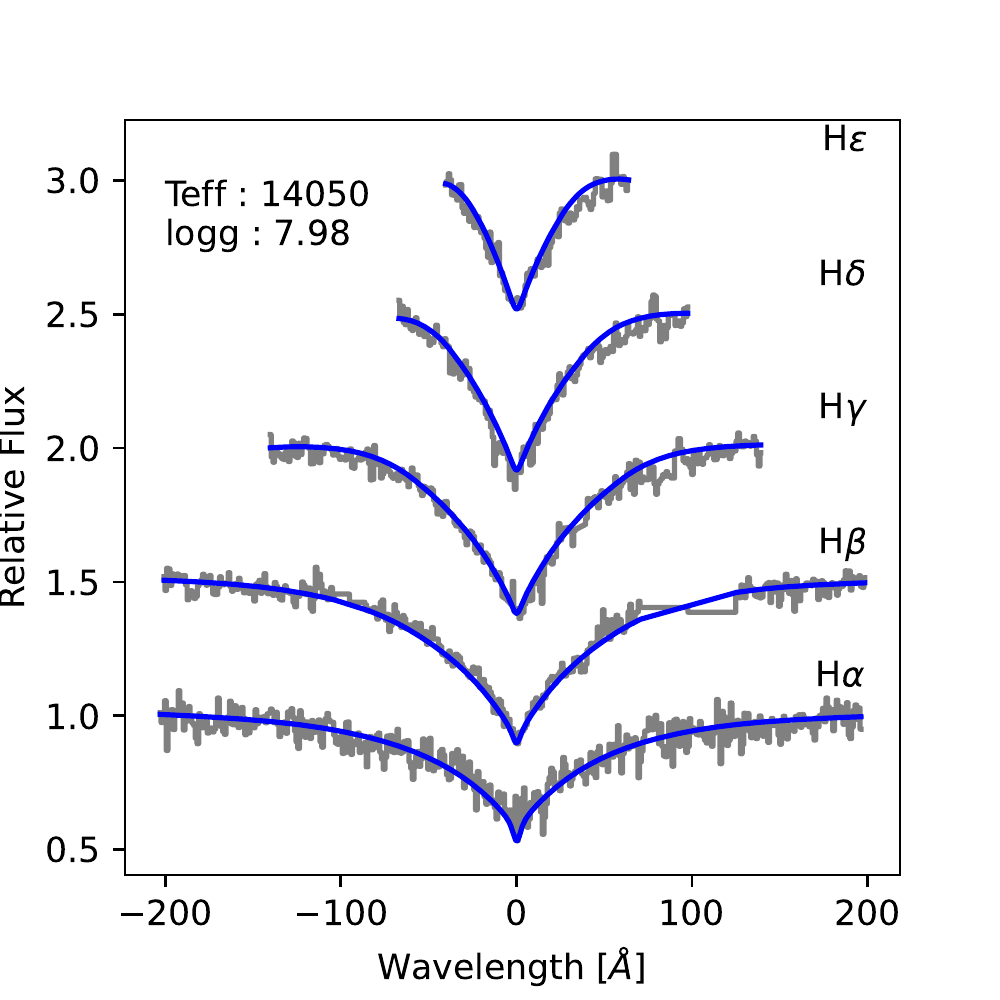}
    \caption{Extracted and normalised Balmer lines from H$\alpha$ (bottom) to H$\epsilon$ (top) are compared to the best-fit model.}
    \label{fig:fitlines}
\end{figure}

We used the blue segment of the spectra to constrain the physical parameters of the white dwarf. At wavelengths shorter than $7000$ \AA, the white dwarf component of TIC 60040774 is the dominant source of light. A set of DA-type white dwarf synthetic spectra with pure hydrogen atmosphere \citep{koester2010} were compared to the observed data using standard procedures \citep{bergeron1992}. Each model atmosphere is parameterised by a certain $\teff{}$ and $\logg{}$. Balmer lines from H$\alpha$ to H$\epsilon$ were extracted and normalised separately. To find the normalisation factor, we fitted a pseudo-Gaussian function to each line profile. DA-type white dwarfs with effective temperatures of $9000\text{ K}\lesssim T_{\text{eff}}\lesssim 16000\text{ K}$ show very strong Balmer lines which are well matched to pseudo-Gaussian functions \citep{bergeron1995}. We implemented the same extraction and normalisation to the synthetic spectra prior to the fitting. Additionally, line profiles from the models were convolved with a Gaussian kernel with FWHM of $2$ \AA\ in order to match the instrumental profile. After some experiments with a coarser parameter grid, $\teff{}$ models ranging from $13000$ to $15000$ K with $250$ K resolution and $\logg{}$ ranging from $7.00$ to $9.00$ with $0.25$ resolution were fitted to the extracted line profiles. The $\chi^2$ was used as the cost function to find the best-fit model and obtain the optimum model parameters. Subsequently, more physical parameters such as mass, radius, and the cooling age of the white dwarf can be estimated using theoretical models \citep{tremblay2013}.

The best fit to the Balmer lines ($H\alpha$ to $H\epsilon$) yields $\teff{,A}=14050\pm370$ K and $\logg{A}=7.98\pm0.07$ (in cgs). Figure \ref{fig:fitlines} displays the extracted line profiles and best fit model. The contamination from the red companion is negligible even at $H\alpha$ wavelength, such that the solution obtained using all available Balmer lines is reliable. Additionally, the deviation between the 1D model and the more realistic 3D atmospheric model at the quoted temperature and gravity is minimal \citep{tremblay2013}. According to the cooling track model provided by the Montreal group \citep{tremblay2011, bedard2020}, the estimated mass of the white dwarf is $M_A=0.598\pm0.029$ M$_{\odot}$, with a corresponding radius of $R_A=0.0131\pm0.0005$ R$_{\oplus}$, while the cooling age is $t_{\text{cool}}=243\pm24$ Myr. The uncertainties for these derived parameters were estimated by randomly sampling the models with possible sets of $\teff{}$ and $\logg{}$ following a distribution dictated by their uncertainties. This solution is consistent with the mass and radius calculated using Koester's model, which was scaled by the observed magnitude in $g$, $B_P$, and $G$ bands. Using bolometric luminosities derived from those magnitudes, the obtained mass and radius are $0.63\pm0.10$ M$_{\odot}$ and $0.0133\pm0.003$ R$_{\odot}$ respectively. This adopts the bolometric correction table from \citet{chen2019} and standard formulae and solar quantities from \citet{prsa2016b}.

With a mass of $0.598$ M$_{\odot}$ and $\teff{}=14050$ K, the primary component of TIC 60040774 can be regarded as a typical white dwarf \citep{kleinman2013}. It is far enough from the instability strip that variability beyond the eclipse is not expected. As the first-order estimate of the system age, the progenitor mass of the white dwarf can be estimated using initial-final mass relations. Its main sequence lifetime could also be estimated using reliable stellar evolution model. In this scheme, the white dwarf is assumed to evolve without significant influence from its companion. Using the semi-empirical relation from \citet{catalan2008}, the estimated initial mass of TIC 60040774a is $1.78\pm0.48$ M$_{\odot}$. Stars with solar metallicities as those inferred here spend $1.49$ Gyr on the main sequence and an additional $0.30$ Gyr during the post-main sequence phase before cooling as white dwarfs \citep{choi2016}. The use of the initial-final mass relationship overestimates the age of TIC 60040774. This is because the mass transfer from a massive progenitor, as well as common envelope evolution, were likely to occur in this system. We will discuss the past evolution of TIC 60040774 Section \ref{sec:evo}.

\begin{figure*}
    \centering
    \includegraphics[scale=0.75]{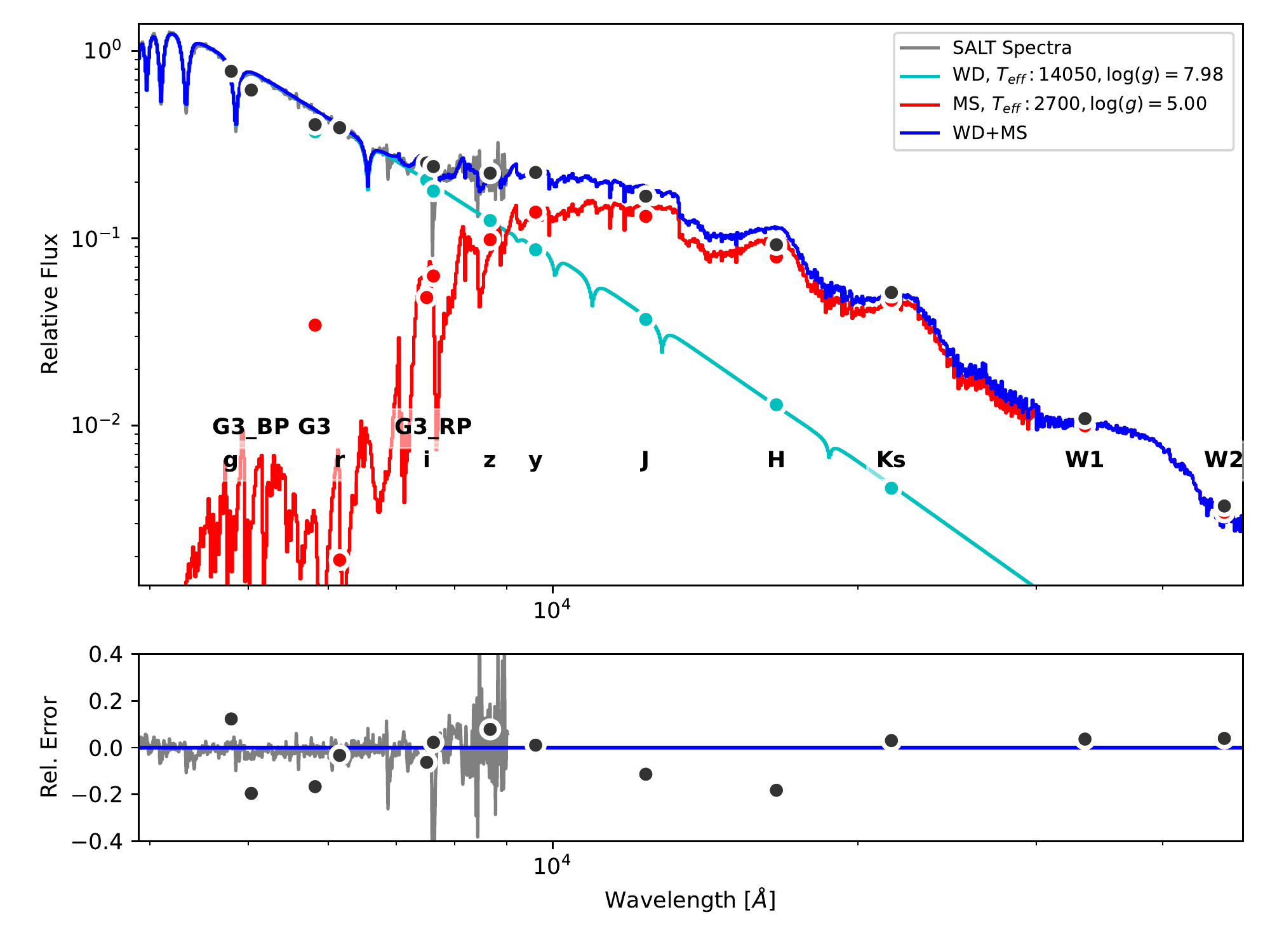}
    \caption{The best-fit model of white dwarf-M dwarf binary (blue) overlaid to the optical spectrum from SALT/RSS (grey) and the observed SED (dark grey). The relative errors (bottom panel) between the model and observation are below $20\%$.}
    \label{fig:fitspec}
\end{figure*}

\subsection{M dwarf parameters}
The flux density of the secondary component ($F_{\lambda,2}$) was estimated by subtracting the observed total flux ($F_{\lambda,\text{total}}$) with the contribution from the white dwarf ($F_{\lambda,1}$) derived from the white dwarf model with $\teff{,A}=14050$ K and $\logg{A}=7.98$. Interpolation over a rectangular mesh of $\teff{}-\logg{}$ was done using a bivariate cubic spline to yield precise synthetic SED for the white dwarf. After that, the fluxes of the secondary component (TIC 60040774B) at specified photometric bands were calculated using the following equation:
\begin{equation}
    F_{\lambda,2} = F_{\lambda,\text{total}}\left[1 -  10^{-0.4(m_{\lambda,1}-m_{\lambda,\text{total}})}\right],
    \label{eq:flux}
\end{equation}
where $m_{\lambda}$ represents the apparent magnitude at a particular band. The result is an SED with a peak at around $1\ \mu\text{m}$, implying that the effective temperature of this star is around $2900$ K.

The SED can also be translated to the corrected magnitudes and colours of the M dwarf. Applying the potometric classification scheme presented by \citet{skrzypek2015} in combination with the empirical relation between colors and spectral types, we estimated that TIC 60040774B is an M6.5 dwarf with an effective temperature of $2600-2900$ K. In this approach, we used the polynomial functions to relate colour indices to the spectral types based on \citet{best2018} The $\chi^2$ minimisation method was based on the prescription from \citet{skrzypek2015}. Colour indices were calculated using magnitudes observed at $i$-band and longer wavelengths. At a shorter wavelengths, the flux from the M dwarf is very low and thus uncertain. The translation from spectral type to temperature is based on \citet{rajpurohit2013}.

In more detail, the calculated SED from the secondary was then compared to the collection of synthetic spectra from BT-Settl CIFIST \citep{caffau2011, allard2012} covering $2000\text{ K}\leq\teff{}\leq4000\text{ K}$ with $100$ K resolution and $2.5\leq\logg{}\leq5.5$ with $0.5$ resolution. BT-Settle CIFIST provides excellent atmospheric models which are compatible with the observed properties of very-low-mass stars and brown dwarfs with solar metallicity. The optimum parameters were obtained through $\chi^2$ minimisation using VOSA online service.

The SED analysis through VOSA resulted in the atmospheric parameters $\teff{,B}=2700\pm150$ K and $\logg{B}=4.14\pm0.97$, where the mean values and the corresponding uncertainties are based on the profile of smoothed $\chi^2$. The quoted uncertainties are associated with the $\chi^2=\chi^2_{\text{min}}+\Delta\chi^2$, where $\Delta\chi^2=2.3$ for two-parameter-model fitting \citep[e.g.,][]{avni1976}. Moreover, VOSA also provided the scaling factor of $M_d=R^2/d^2=(5.54\pm0.60)\times10^{-22}$, which can be used to derive the radius of the object, which is $R_{2}=0.140\pm0.008$ R$_{\odot}$. From the adopted surface gravity and radius, the derived mass of the dwarf is $\log(M/M_{\odot})=-1.98\pm0.98$. The estimated radius is in good agreement with the one calculated using the semi-empirical relation between radius and absolute magnitude in the $Ks$ band (calibrated using nearby K7--M7 dwarfs \citep{mann2015}). Using the corrected magnitude and the distance from \textit{Gaia} EDR3, we got $M_{Ks}=9.26\pm0.10$ and a radius of $0.133\pm0.004$. Additionally, based on the mass-luminosity relations from \citet{mann2019}, we estimated that the secondary component has a mass of $0.107\pm0.020$ M$_{\odot}$. Though the mass-luminosity relation from \citet{mann2019} is more accurate, especially for low-mass stars, the intrinsic scatter in the relation becomes the major source of the quoted uncertainty. The latter mass estimate is more plausible than the gravity-based estimate since the uncertainty in the $\logg{}$ from the SED fitting is large.

Figure \ref{fig:fitspec} shows the comparison between the observed SED with the model containing $14050$ K white dwarf with $\logg{B}=7.98$ and the M dwarf with $\teff{,B}=2700$ K and $\logg{B}=5.0$. The adopted surface gravity of the secondary is higher than the solution from VOSA to comply with the mass-luminosity relations mentioned before.

\subsection{Light curve modelling}
The folded and binned light curve was modelled using \textsc{phoebe2} \citep{prsa2016}. This code is the updated version of \textsc{phoebe} \citep{prsa2005}, which includes several improvements, especially the inclusion of the library of stellar atmospheric models, various observing passbands, and also valuable tools for inverse problem solving \citep{horvat2018, jones2020, conroy2020}. To model the observable quantities as a function of time or phase, the code creates a gravitationally affected surface mesh of model stars. It populates the triangular mesh with quantities such as effective temperature, intensity, and colour based on relevant physics. For a WDMS binary, the white dwarf can be represented as a blackbody radiator, while the atmosphere of the cooler star can be calculated using PHOENIX, covering $\teff{}$ down to $2300$ K \citep{husser2013}. The \textit{TESS} passband covering $6000-10000$ \AA\ was used to compute the expected variability more accurately. Limb darkening of the primary component was represented by a power-law model with appropriate coefficients from \citet{claret2020}. For the secondary component, we used a quadratic limb darkening function with appropriate coefficients from \citet{claret2018}. Estimated $\teff{}$ and $\logg{}$ become the parameters for selecting the limb darkening coefficients.

The geometry of the folded light curve, especially the timing of ingress and egress and the eclipse depth, provides constraints on some of the stellar properties. Those include the radii of primary and secondary scaled by the semimajor axis and the ratio of the effective temperatures of both components. However, there is known degeneracy of the solutions due to the orbital inclination. For example, a model with larger stellar radii and lower inclination may produce a similar light curve profile as the model with smaller radii and higher inclination. There are some ways to relieve this degeneracy \citep{parsons2017}, but those approaches require precise multi-epoch spectroscopic data to characterise the kinematics of the components. Consequently, light curve modelling must depend on prior knowledge established by fitting synthetic spectra and SEDs. Nonetheless, light curve modelling can provide additional consistency checks on parameters constrained by previous methods.

A Markov Chain Monte Carlo (MCMC) sampling was performed to obtain posterior probability distributions of parameters after the light curve modeling. \textsc{phoebe2} includes this capability by incorporating Affine-invariant MCMC algorithms \citep{goodman2010} implemented in the \textsc{emcee} Python package \citep{foreman2013, foreman2019}. In this approach, the light curve models are governed by five free parameters sampled from the initial parameter distributions. These are the orbital inclination $i$, equivalent radii ($R$), and effective temperature ($\teff{}$) of primary and secondary components. The orbital eccentricity was set to zero while the period was fixed. The masses of the components cannot be evaluated directly using the light curve, but the masses constrain the semimajor axis, which in turn scales the radii. We established the initial distribution of those parameters based on the results of spectral and SED fitting. To achieve reliable posteriors, we ran MCMC with 10 walkers through 5000 iterations. During that run, we used binary models with 400 triangular mesh.

Table \ref{tab:mcmc} summarises the way we generated the distribution of the priors. The orbital period was fixed at $0.404807$ days, while the circular orbit was assumed since there is no feature in the observed light curve to constrain the eccentricity. If the size ratio between the primary and secondary is $R_B/R_A\approx10$ and the effective temperature ratio is $\teff{,B}/\teff{,A}\approx0.2$, then the expected reduction in bolometric flux is less than two parts-per-thousand. With the $\tmag\approx17$, the photometric error of \textit{TESS} is larger that the secondary eclipse. Besides that, the out-of-eclipse light curve is relatively flat, implying that the irradiation or reflection is insignificant. The eclipse timing might not be exact, but initial sampling experimentation with varied eclipse epoch $T_0$ revealed that the optimum solution deviates by a negligible ${\lesssim}10^{-5}$ days. Consequently, $T_0$ was not included as a free parameters during the final run.

\begin{table}
    \caption{Distribution of priors for MCMC random sampling.}
    \label{tab:mcmc}
    \centering
    \begin{tabular}{l|l|l}
    \hline
    Parameter & Distribution & Unit\\
    \hline
    $\teff{,A}$ & $\mathcal{N}(14050, 370)$ & K\\
    $\teff{,B}$ & $\mathcal{N}(2700, 150)$ & K\\
    $R_{\text{equiv},1}$ & $\mathcal{N}(0.0131, 0.0005)$ & R$_{\odot}$\\
    $R_{\text{equiv},2}$ & $\mathcal{N}(0.1330, 0.0080)$ & R$_{\odot}$\\
    $i$ & $\mathcal{U}(85, 90)$ & degree\\
    \hline
    $a$ & $2.0405$ & AU \\
    $P$ & $0.404807$ & day \\
    \hline
    \end{tabular}
\end{table}

\begin{table}
    \caption{Orbital parameters derived from the light curve of TIC 60040774.}
    \label{tab:orbit}
    \centering
    \begin{tabular}{l|l|l}
    \hline
    Period & $P$ & $0.404807\pm0.000149$ days \\
    Time of eclipse & $T_{0}$ & $2459201.762996$ BJD \\
    Eclipse duration & $T_{14}$ & $11.19$ min \\
    Eclipse depth & & $77.87\%$ \\
    Semimajor axis & $a$ & $2.041\pm0.039$ R$_{\odot}$ \\
    Inclination & $i$ & $87.50\pm0.16^{\circ}$ \\
    \hline
    \end{tabular}
\end{table}

\begin{table*}
    \caption{Stellar properties derived from spectroscopic and photometric data with remarks on the methods.}
    \label{tab:stellar}
    \centering
    \begin{tabular}{l|l|l|l}
    \multicolumn{4}{l}{\textbf{Primary Component}}\\
    \hline
    Mass & $M_A$ & $0.598\pm0.029$ & spectroscopy, cooling track \\
    Radius & $R_A$ & $0.0131\pm0.0005$ R$_{\odot}$ & spectroscopy, cooling track (LC prior) \\
    && $0.0137\pm0.0005$ R$_{\odot}$ & spectroscopy, light curve (LC posterior) \\
    Roche-lobe filling factor & & $1.3\%$ & photometry, light curve\\
    Effective temperature & $T_{\text{eff},A}$ & $14050\pm360$ K & spectroscopy, light curve \\
    Surface gravity & $\log(g_A)$ & $7.98\pm0.07$ & spectroscopy \\
    Cooling age & $t_{\text{cool}}$ & $0.243\pm0.024$ Gyr & spectroscopy, cooling track \\
    \hline
    \\
    \hline
    \multicolumn{4}{l}{\textbf{Secondary Component}}\\
    Mass & $M_B$ & $0.107\pm0.020$ & photometry, mass-luminosity relation \\
    Radius & $R_B$ & $0.1330\pm0.0080$ R$_{\odot}$ & photometry, radius-luminosity relation (LC prior)\\
    && $0.1385\pm0.0040$ R$_{\odot}$ & photometry, light curve (LC posterior)\\
    Roche-lobe filling factor & & $28.5\%$ & photometry, light curve\\
    Effective temperature & $T_{\text{eff},B}$ & $2700\pm150$ K & photometry, SED fitting (LC prior)\\
    && $2759\pm50$ K &photometry, light curve (LC posterior)\\
    Surface gravity & $\log(g_B)$ & $5.00\pm0.50$ & photometry, SED fitting\\
    Spectral class & & M6.5 & photometry\\
    \hline
    \end{tabular}
\end{table*}

\begin{figure*}
    \centering
    \includegraphics[width=0.8\textwidth]{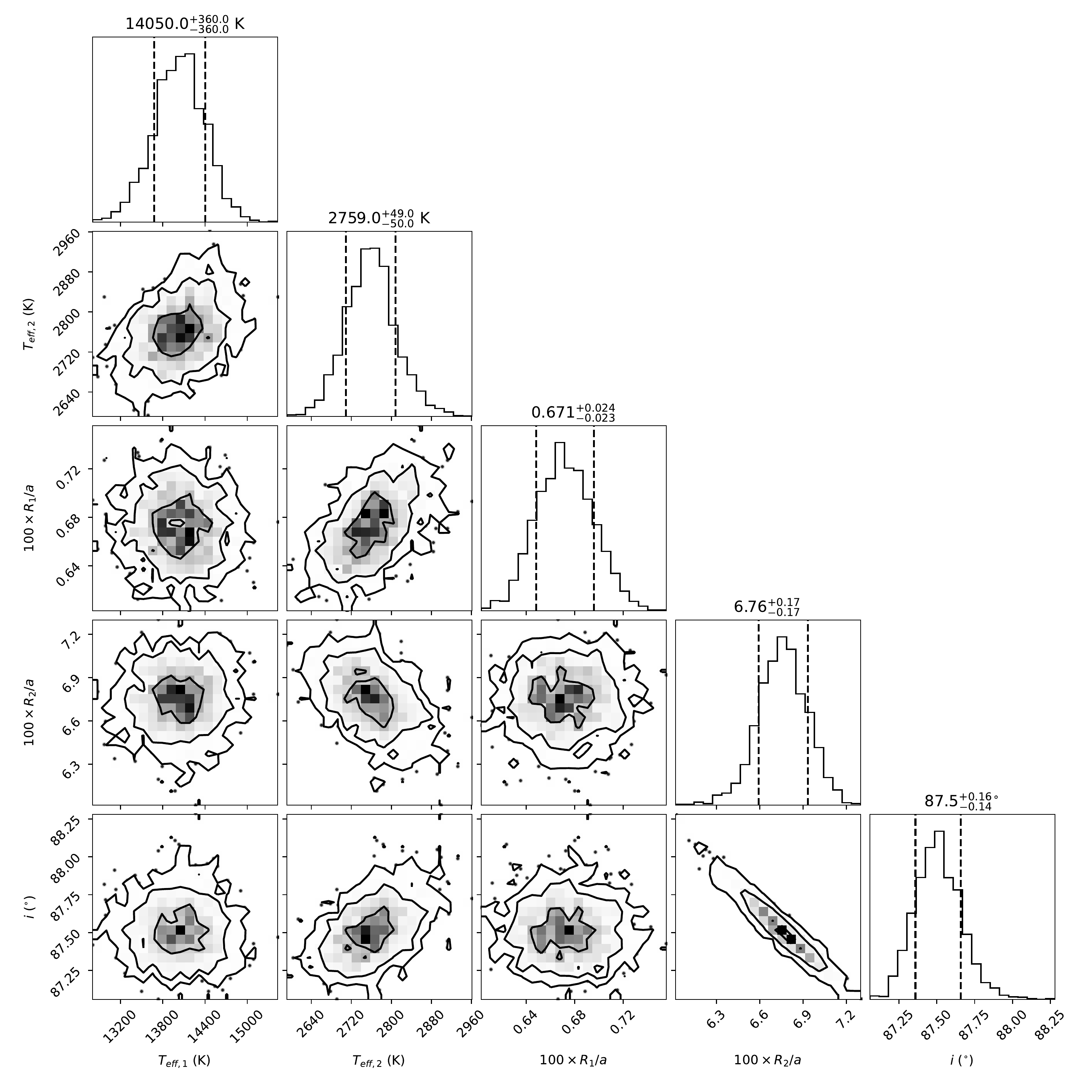}
    \caption{Corner plot showing the posterior distributions of some parameters of TIC 60040774.}
    \label{fig:emcee}
\end{figure*}

Figure \ref{fig:emcee} displays the posterior distribution obtained from the MCMC run. From that figure, some parameters correlate with each other. The highest correlation is between the inclination and the radius of the secondary. The solution of $R_B/a=0.0676$ and $i=87.5^{\circ}$ achieved through the MCMC sampling is consistent with the solution from the SED fitting. The depth of the light curve determines the radius of the secondary such that the posterior distribution is more constrained than the prior (improvement from $6\%$ to $2\%$ relative error). This also applies to the secondary's effective temperature even though weak positive correlations exist between $\teff{,B}$ and other parameters ($i$ and $R_{A}$). On the other hand, the posterior distributions for $\teff{,A}$ and $R_A$ are not different to the prior distributions, implying that the single band light curve provides insufficient information to constrain the parameters of the primary component. Lastly, the light curve cannot provide additional information to constrain the masses. Further spectroscopic observation that produces a sufficient radial velocity profile is required to constrain the masses of the components.

\begin{figure}
    \centering
    \includegraphics[width=\columnwidth]{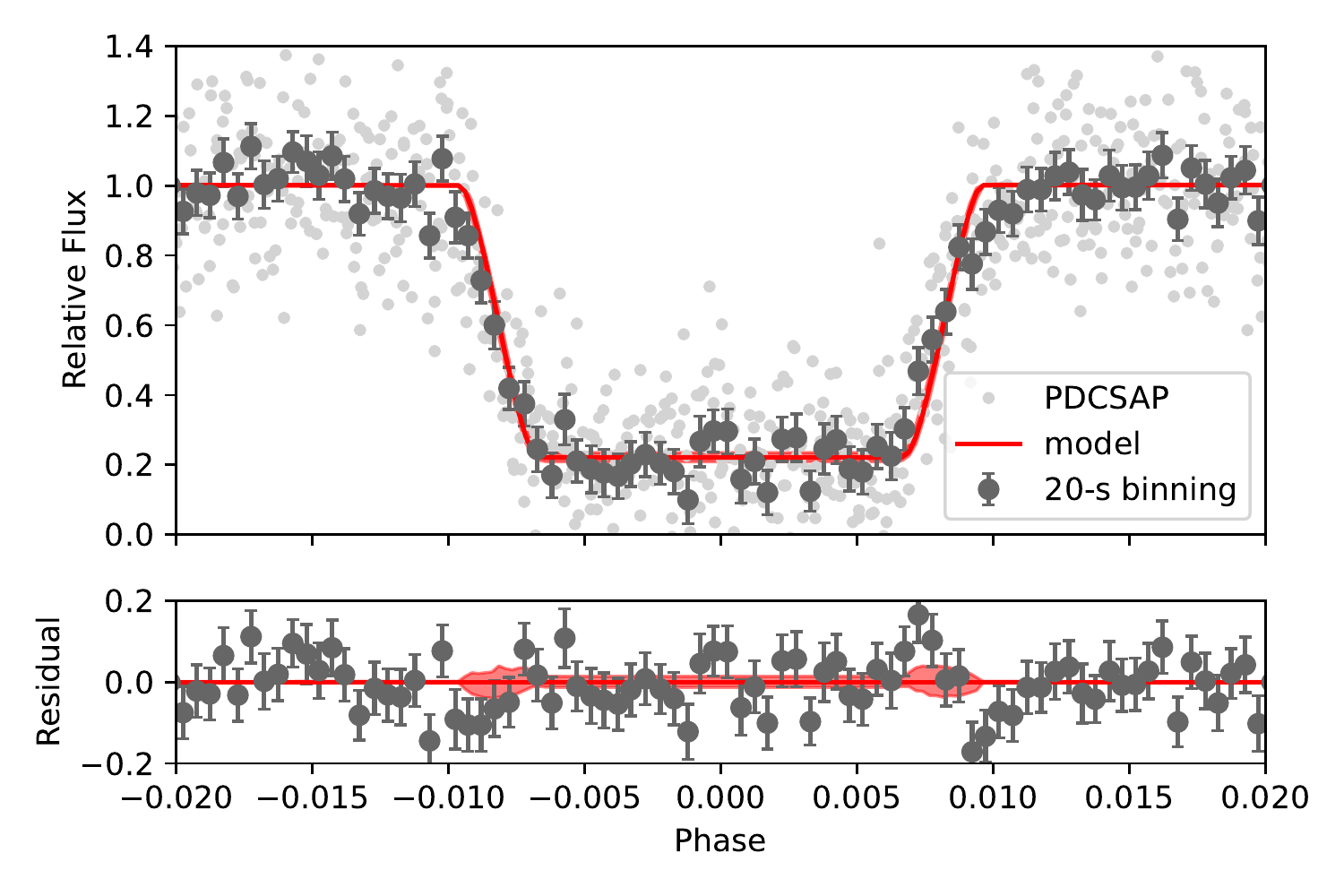}
    \caption{The best fit eclipse model and its $1\sigma$ uncertainty for the primary eclipse of TIC 60040774 as observed by TESS. Circles with error bars are 20-s binned data.}
    \label{fig:eclipse}
\end{figure}

\begin{figure*}
    \centering
    \includegraphics[width=0.8\textwidth]{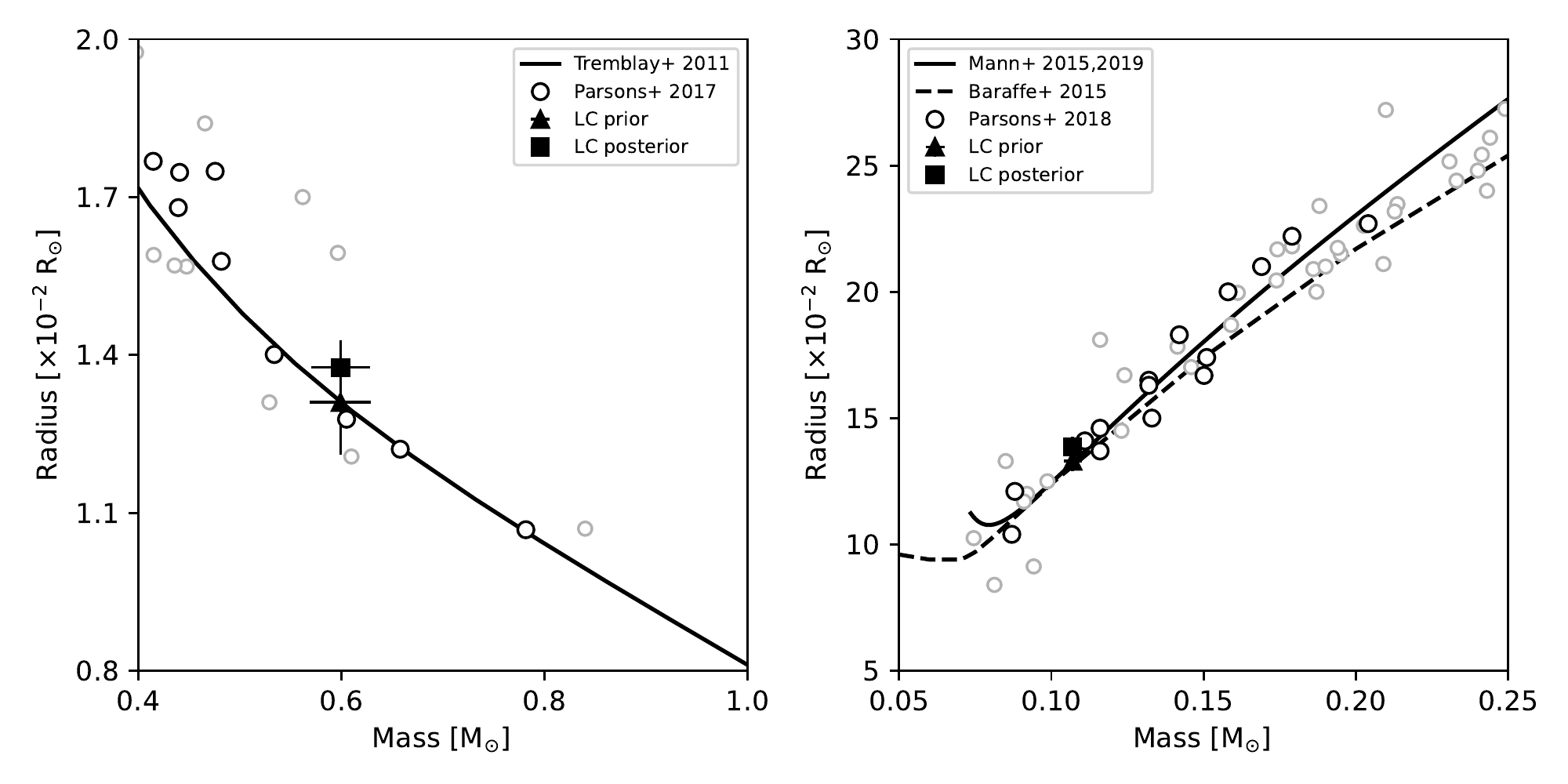}
    \caption{Mass-radius plot for the white dwarf primary (left) and secondary (right) components of TIC 60040774 based on the solutions prior (triangle) and after (square) light curve modeling. For the white dwarf, the plot is complemented with white dwarf sample from \citet{parsons2017} and theoretical relation from \citep{tremblay2011}. Darker circles are for white dwarfs with $10,000\leq T_{\text{eff}}\leq20,000$ K. For the secondary M-dwarf, samples from \citep{parsons2018} are overplotted, together with theoretical model from \citet{baraffe2015} and semi-empirical relations from \citet{mann2015} and \citet{mann2019}}.
    \label{fig:massradius}
\end{figure*}

The best-fit light curve model with uncertainty is displayed in Figure \ref{fig:eclipse}, while the parameters are summarised in Table \ref{tab:orbit} and Table \ref{tab:stellar}. For this model, the goodness of fit is represented by the reduced $\chi^2$ of $1.05$. Based on the adopted masses, we estimated that the Roche-lobe filling factor of the primary and secondary are $1.3\%$ and $28.5\%$ respectively. This system is fully detached with nearly spherical components. The posterior parameters are also plotted in Figure \ref{fig:massradius}. For the primary, it is seen that the MCMC run provides a radius $5\%$ larger than the one derived from the cooling track.

As we can see in Figure \ref{fig:eclipse}, the ingress and egress profile are shallower than the best fit model, implying that there is a possibility that the geometry of either primary or secondary is not so simple. To explore this further we experimented with different limb darkening profiles (quadratic and square-root). However we find that different profiles cannot explain the shallower ingress-egress without inflating the white dwarf radius. Shallower ingress-egress may be produced by a bright accretion disc normally observed in cataclysmic variables \citep[e.g.,][]{savoury2011}. If this is the case, then the light curve morphology implies that the disc extends to ${\gtrsim}5R_A$ and contributes ${\gtrsim}10\%$ to the observed luminosity. Nevertheless, this explanation is not suitable for TIC 60040774 where the secondary is far from filling its Roche-lobe. Additionally, the spectrum of this object does not show any hydrogen emission.

Alternatively, a planetary debris disc may exist around the primary since the chance of finding a young white dwarf (up to ${\sim}200$ Myr cooling age) with planetary debris is ${\sim}27\%$ or higher \citep{koester2014}. In most cases, the identification and characterisation of planetary debris discs around white dwarfs are based on infrared observations. The typical outer radius of the disc is $0.3$ R$_{\odot}$ \citep{farihi2016}. Transiting planetary debris in front of a white dwarf produces shallower and variable transit depths \citep{vanderburg2015, vanderbosch2021}. The asymmetric profile of ingress and egress may also lead to the notion that the secondary is surrounded by debris, with a trailing part that is thicker than the leading part. Alternatively, an extended atmosphere of the secondary may also produce the shallower ingress-egress by gradually absorbing the radiation from the compact primary \citep{day1988}. Apart from these alternative explanations, there is still a possibility that the observed profile comes from systematic error. Consequently, follow-up observations with high cadence photometry covering the complete eclipse stages are needed to clarify the true ingress-egress profile of TIC 60040774.

\subsection{Binary evolution of TIC 60040774}
\label{sec:evo}
The white dwarf-main sequence binary TIC 60040774 is expected to have formed as a system with higher total mass and wider orbit than those currently observed. The system experienced binary evolution that includes a common-envelope (CE) phase. This phase lasted for a very short time but significantly shrank the orbit by orders of magnitude. At the end of the CE phase, the orbit shrank further, mainly due to gravitational radiation. Inversely, we can trace possible configuration of TIC 60040774 in the past from the observed parameters.

First, the system's orbital period just after the CE phase ($P_{CE}$) can be estimated by considering the angular momentum loss due to gravitational radiation. For the case of TIC 60040774 where the secondary is a fully-convective star with mass $m_2<0.35$ M$_{\odot}$, magnetic braking is inefficient in shrinking the orbit. Following \citet{schreiber2003},
\begin{equation}
    P_{CE}^{8/3} = \dfrac{256G^{5/3}(2\pi)^{8/3}}{5c^5}\dfrac{M_AM_B}{M_A+M_B}t_{\text{cool}} + P_{\text{orb}}^{8/3}
\end{equation}
where $G$ is the gravitational constant, $c$ is the speed of light in a vacuum, $M_A$ and $M_B$ are the current masses of the primary and secondary components, $P_{\text{orb}}$ is the observed orbital period, while the $t_{\text{cool}}$ is the white dwarf cooling age. Using the parameters presented in Table \ref{tab:orbit} and Table \ref{tab:stellar}, we got $P_{CE}=0.40504$ days.

We performed forward modelling using the Binary Star Evolution (BSE) code \citep{hurley2002} which includes single star evolution recipe \citep{hurley2000} and various physical mechanisms involved in the interacting binary star evolution. Synthetic binaries with a fixed secondary mass of $m_2=0.107$ M$_{\odot}$ and various progenitor masses of $1.80\leq m_1\leq3.80$ M$_{\odot}$ were generated and evolved until they reached the common-envelope phase. Those binaries have circular orbits with initial periods of $400\leq P_{\text{init}}\leq1000$ days or separations between $1.33$ to $3.07$ AU. The resolution of input parameter grids is $0.01$ M$_{\odot}$ in mass and $5$ days in period. We adopted solar metallicity and set the remaining stellar evolution parameters to default values. We used the common-envelope efficiency $\alpha_{CE}=1.0$ and the binding energy factor $\lambda=0.5$. These two parameters have a significant effect on the outcome of the common-envelope evolution. According to \cite{davis2012}, the $\alpha_{CE}$ correlates with the observed white dwarf mass and secondary component. The adopted values for the BSE models are consistent with the prescribed relation assuming that the binding energy of the envelope is purely gravitational. The adopted $\lambda$ is based on the calculation by \cite{dewi2000} for the star of mass 3 M$_{\odot}$.

The simulation produced some important parameters at the end of the CE phase, especially the orbital period ($P_{CE}$) and the mass of the white dwarf to be compared to the ones derived from the observation. The best estimate of the progenitor mass is $m_1=2.59\pm0.33$ M$_{\odot}$ while the initial period and separation are $829\pm115$ days and $2.40\pm0.24$ AU. When the primary started to reach the giant branch, the wind from the primary induced drag and shrank the orbit to the point of Roche-lobe overflow. This process was followed by the CE phase.

Beyond the common-envelope stage, the evolution of TIC 60040774 is straightforward without any dynamical complexity. The white dwarf is getting cooler while the low-mass secondary evolves slowly, and it will not fill its Roche-lobe in the next couple of billion years. Gravitational-wave radiation with a time scale of 400 Gyr becomes the only mechanism reducing the orbital separation unless the envelope ejection is not 100\% efficient ($\alpha_{CE}\ll1$) and the existence of a debris disc accelerates the orbital shrinkage \citep{kashi2011}.


\subsection{TIC 60040774 among other eclipsing PCEBs}
Having a primary mass of $0.598$ M$_{\odot}$, a secondary mass of $0.107$ M$_{\odot}$, and orbital period of $9.71$ hours, TIC 60040774 can be regarded as a typical WDMS binary as it is located close to the central distribution of WDMS in terms of white dwarf mass, temperature of the secondary component, and also the orbital period \citep{rebassa2016, rebassa2021, kruckow2021}. However, TIC 60040774 gains its uniqueness if we place it in the context of eclipsing PCEB.

\begin{figure}
    \centering
    \includegraphics[width=\columnwidth]{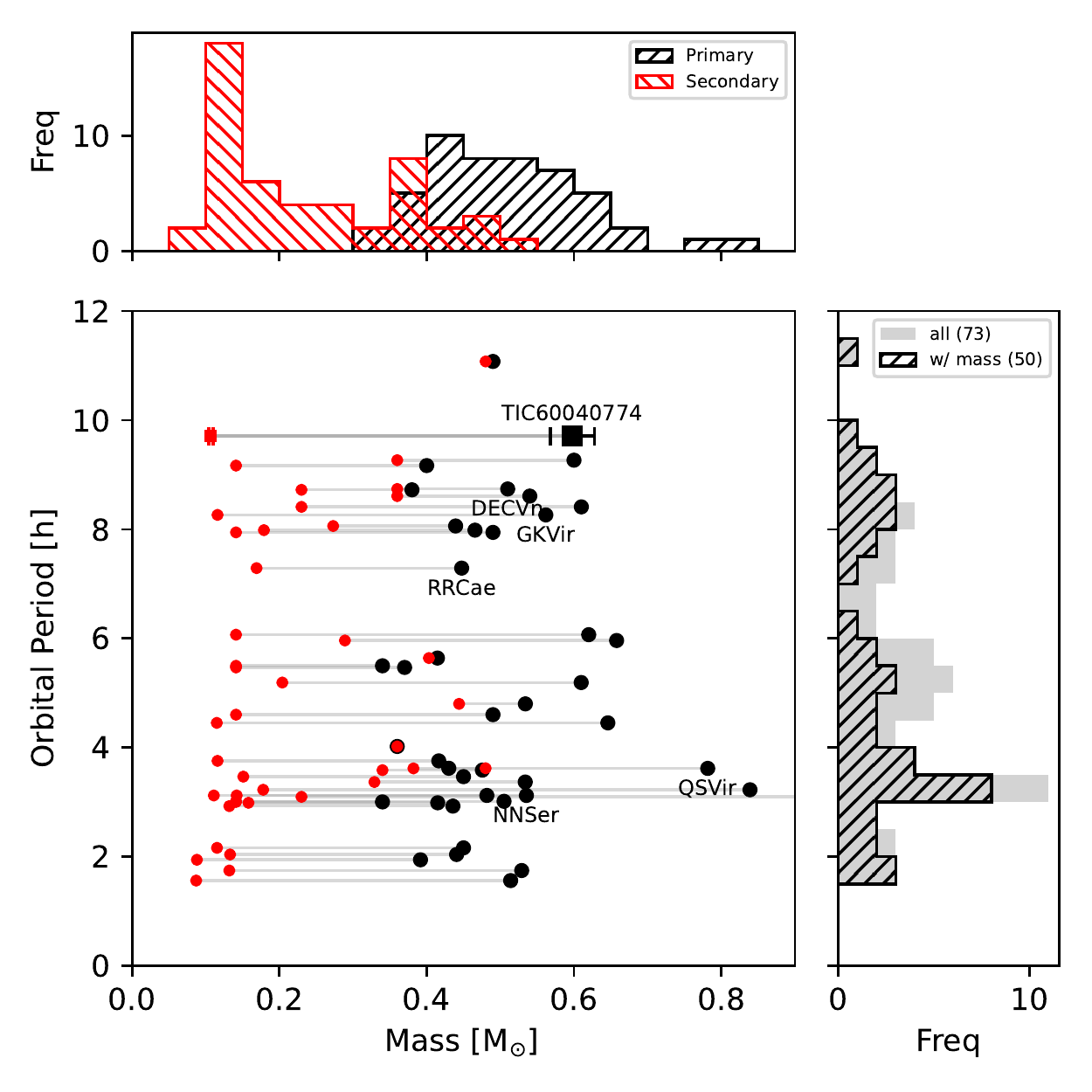}
    \caption{Plot of the orbital periods and masses of 40 eclipsing PCEBs listed in \citep{parsons2015} where the primary and secondary masses are in black and red respectively. The associated histograms are displayed where the gray histogram is for the complete sample.}
    \label{fig:massporb}
\end{figure}

Figure \ref{fig:massporb} shows the distribution of mass and period of PCEBs listed in \citet{parsons2015} and \citet{parsons2017}. The distribution of the primary mass shows a single peak at around $0.4-0.45$ M$_{\odot}$, while the distribution of secondary mass is double-peaked with a clear valley around $0.35$ M$_{\odot}$. This valley is associated with the upper limit of fully-convective stars. Systems with such low-mass secondaries remain detached for an extended period since the orbital shrinkage is only caused by gravitational wave radiation \citep{zorotovic2014}. Consequently, the probability of finding such systems is higher. TIC 60040774 is among the systems with a low-mass secondary.

If we look at the period distribution, eclipsing PCEBs tend to have a shorter orbital period while TIC~60040774 is at the right-end of the distribution. We can say that TIC~60040774 is a relatively rare case because the probability of finding eclipsing system is inversely proportional to the orbital separation (and period). Besides this trend, a multimodal distribution is seen in Figure \ref{fig:massporb}. However, this distribution is likely to be caused by selection bias as the Monte Carlo simulation of the PCEBs \citep{zorotovic2014} did not produce the multimodality.

With the orbital separation of ${\sim}2$ AU, TIC 60040774 is also a simple detached system where the irradiation from the primary star does not affect the secondary. The effective temperature of the white dwarf is around five times that of the secondary, while the scaled radius of this compact body is just $0.00671$. This condition implies that the irradiation from the white dwarf is just $1.5\%$ that of the intrinsic flux of the M-dwarf secondary. Consequently, the SED of the secondary is close to the theoretical model.

Eclipsing PCEBs with low-mass secondary become a good target for finding circumbinary planets since the sharpness of the light curve implies a great precision of measuring the eclipse timing. Systems like NN Ser, QS Vir, RR Cae, GK Vir, and DE CVn are known to have circumbinary planet(s) with orbital period of more than ten years \citep{beuerman2010,qian2012,han2018,almeida2020}. The existence of a Jupiter-like planets modulates the orbital period of the binary with typical amplitudes of a few tens of seconds by inducing light time travel effects \citep[e.g.,][]{marsh2014}. Either the first generation formation scenario \citep{bear2014} or the post-common-envelope planetary formation \citep{schleicher2014} may explain the existence of planets orbiting PCEBs. Belongs to the same class, TIC 60040774 may also host a planet or planetary system. However, \citet{bear2014} speculated that the absence of detection of a planet orbiting PCEB with ${\sim}0.6$ M$_{\odot}$ primary may be a consequence of evolutionary process. At this moment, we have not identified any extra transit in the \textit{TESS} light curve, although this only covers one-month of observations. Even though there are more than 60 eclipses observed for TIC 60040774, the unbinned light curve from a single eclipse is noisy such that identifying short period modulation from this data is difficult. Similar to the case of GK Vir \citep{almeida2020}, follow-up observations using a meter-class telescope for more than a decade is required to reveal the existence of a planet orbiting TIC 60040774.

\section{Conclusion}
We derived orbital and physical parameters of TIC 60040774, one of the eclipsing post common-envelope binary with low-mass secondary located $134$ pc from the Sun. Based on spectroscopic data, we estimated that the primary component of this system is a white dwarf with a mass of $0.598\pm0.029$ M$_{\odot}$ and an effective temperature of $14050\pm360$ K. The secondary component is an M6.5 dwarf with a mass of $0.107\pm0.020$ M$_{\odot}$ and a temperature of $2759\pm50$ K. Physical parameters for the secondary were derived through SED fitting combined with the MCMC sampling to fit the light curve from TESS. With the period of 9.71 hours, TIC 60040774 becomes one of the eclipsing PCEBs with relatively long orbital period. Additionally, we estimated that the system started from binary with a late B primary star of mass $2.5$ M$_{\odot}$ and an initial separation of $2.4$ AU. The total age of the system is $1.024\pm0.300$ Gyr with the common-envelope stage occuring $0.243$ Gyr ago.

Our estimate of the parameters of the TIC 60040774 is not independent of some theoretical and semi-empirical models. However, the orbital separation is relatively large such that the irradiation from the white dwarf does not affect the secondary. Consequently, the expected properties are consistent to the models. However, we found that the ingress-egress profile acquired from the \textit{TESS} photometry seems to be shallower than the one expected from the best model consisting of components with typical geometrical structure. This profiles may indicate the existence of debris disc in TIC 60040774 though this hypothesis needs to be evaluated using additional data. 

\section*{Acknowledgment}
This paper includes data collected by the \emph{TESS} mission. Funding for the \emph{TESS} mission is provided by the NASA's Science Mission Directorate. This work has made use of data from the European Space Agency (ESA) mission {\it Gaia} (\url{https://www.cosmos.esa.int/gaia}), processed by the {\it Gaia} Data Processing and Analysis Consortium (DPAC, \url{https://www.cosmos.esa.int/web/gaia/dpac/consortium}). Funding for the DPAC has been provided by national institutions, in particular the institutions participating in the {\it Gaia} Multilateral Agreement. This publication makes use of VOSA, developed under the Spanish Virtual Observatory project supported by the Spanish MINECO through grant AyA2017-84089. VOSA has been partially updated by using funding from the European Union's Horizon 2020 Research and Innovation Programme, under Grant Agreement no 776403 (EXOPLANETS-A). The SALT observations were obtained under the SALT transient followup programme 2018-2-LSP-001 (PI: DAHB). DAHB and JB acknowledge reseach support from the National Research Foundation.

\section*{Data Availability}
The data underlying this article will be shared on reasonable request to the corresponding author.

\bibliographystyle{mnras}
\bibliography{main}

\label{lastpage}

\end{document}